\documentclass[journal]{IEEEtran}
\usepackage{graphicx}
\usepackage{multirow}
\usepackage{hyperref}
\usepackage{url}
\usepackage{tabularx}

\title{The Need for Advanced Intelligence in NFV Management and Orchestration}

\author{Dimitrios Michael~Manias and
        Abdallah~Shami \thanks{Dimitrios Michael Manias and Abdallah Shami are with the Department of Electrical and Computer Engineering at the University of Western Ontario e-mail: \{dmanias3, Abdallah.Shami\}  @uwo.ca}}

\begin{document}

\maketitle

% As a general rule, do not put math, special symbols or citations
% in the abstract or keywords.
\begin{abstract}
With the constant demand for connectivity at an all-time high, Network Service Providers (NSPs) are required to optimize their networks to cope with rising capital and operational expenditures required to meet the growing connectivity demand. A solution to this challenge was presented through Network Function Virtualization (NFV). As network complexity increases and futuristic networks take shape, NSPs are required to incorporate an increasing amount of operational efficiency into their NFV-enabled networks. One such technique is Machine Learning (ML), which has been applied to various entities in NFV-enabled networks, most notably in the NFV Orchestrator. While traditional ML provides tremendous operational efficiencies, including real-time and high-volume data processing, challenges such as privacy, security, scalability, transferability, and concept drift hinder its widespread implementation. Through the adoption of Advanced Intelligence techniques such as Reinforcement Learning and Federated Learning, NSPs can leverage the benefits of traditional ML while simultaneously addressing the major challenges traditionally associated with it. This work presents the benefits of adopting these advanced techniques, provides a list of potential use cases and research topics, and proposes a bottom-up micro-functionality approach to applying these methods of Advanced Intelligence to NFV Management and Orchestration.

\end{abstract}

% Note that keywords are not normally used for peerreview papers.

\section{Introduction}
In 2012 the European Telecommunications Standards Institute proposed the concept of Network Function Virtualization (NFV) \cite{AA1}. NFV was initially conceptualized to address challenges faced by Network Service Providers (NSPs) worldwide. With increasing connectivity demands and the increase in network requirements for emerging technologies (\textit {i.e.}, Industrial Internet of Things (IIoT), Intelligent Transportation Systems (ITS), and wearables), NSPs were tasked with expanding their networks to accommodate the growth in demand and improving network performance. This expansion, if conducted, would have accumulated massive amounts of capital and operational expenditures as each network function had its dedicated piece of hardware, and networks were based on ridged physical infrastructure. With the introduction of NFV technology, network functions were abstracted from their dedicated hardware and executed as software-based Virtual Network Functions (VNFs) in servers, data centers, and network clouds. \par 
While NFV technology can potentially lead to reduced capital and operational expenditures and improved network health (\textit {i.e.}, portability, scalability, availability) \cite{BB1}, several challenges are yet to be addressed regarding the Management and Orchestration (MANO) of these functions. The main challenges related to NFV MANO include placement, chaining, scaling, migration, fault recovery, and security. As new technologies are introduced, new challenges that must be addressed by NSPs arise. Currently, with the incoming introduction of Fifth Generation (5G) networks, a new set of challenges relating to intelligent MANO, network automation, and increased functionality have been identified for the use of NFV technology in 5G networks \cite{CC1}. \par
With the constant increase in network connectivity demand, network traffic and patterns are continually changing. This change poses a very significant challenge for NSPs regarding NFV MANO. The benefit of NFV technology is maximized when VNFs are strategically placed through the network to provide a service to the end-user. Service Function Chains (SFCs) are a group of interconnected and interdependent VNF instances that must be traversed to provide the functionality to an end-user. When considering the placement of VNF instances, there are several considerations related to Quality of Service (QoS) guarantees and Service Level Agreements (SLAs) that go into selecting an appropriate placement. QoS guarantees and SLAs outline various requirements regarding network health (i.e., performance, availability, and reliability), and the maximization of their associated metrics is one of the main challenges for NSPs as the dynamic nature of the networks makes this inherently more challenging. For more information on NFV and NFV MANO, we refer the readers to the following comprehensive sources \cite{RR1}, \cite{RR2}, \cite{RR3}. \par
Traditionally, the problem of the initial VNF Placement and any migration and scaling actions have been formulated as optimization problems due to their NP-Hard complexity \cite{DD1}. However, since dynamic networks require real-time decisions, optimization problem formulations are rendered ineffective due to their time-complexity. As an alternative, several near-optimal heuristic solutions have been proposed to address the NP-Hard complexity of these problems and to improve runtime efficiency. While more time-efficient, heuristic solutions still do not meet the real-time requirements of NFV-enabled networks. Additionally, the near optimality of the heuristic solutions is an undesirable aspect as network performance is not optimized, and therefore financial losses are incurred. Recently, the use of Machine Learning (ML) has been used to address the NP-Hard complexity of the VNF Placement problem. Using ML, intelligence is injected into the network in a time-efficient manner. Additionally, by using previously calculated optimal placements and decision, the ML model can learn and emulate optimality in real-time. However, considering the profound impact traditional ML has had on NFV MANO functionalities, it fails to address the dynamic nature of networks; Traditional ML models are trained on a dataset, however, with network topologies and configurations constantly changing, the dataset must also change accordingly, something which is very time consuming and infeasible. In terms of datasets, an additional challenge faced by NSPs is the privacy concerns surrounding user data. To address the challenges faced by traditional ML modeling in dynamic network environments, next-generation ML, known as Advanced Intelligence, is being proposed as a solution. The remainder of the paper is organized as follows. Section 2 outlines the motivation behind using Advanced Intelligence. Section 3 outlines Reinforcement Learning along with its benefits and use cases. Similarly, Section 4 outlines the Federated Learning and its associated benefits and use cases. Section 5 discusses the next steps. Finally, Section 6 summarizes and concludes the paper.

\section{Motivation}
Advanced Intelligence is a moniker for high performing adaptive intelligence and new intelligence technology. Two such intelligence methods, Reinforcement Learning (RL) and Federated Learning (FL), will be explored in further detail in this paper. However, before explaining what RL and FL are, it is important to establish the need for such forms of intelligence. The main challenges associated with the implementation of ML in NFV MANO will be highlighted to properly illustrate the motivation behind the use of Advanced Intelligence.

\subsection{Increasing Complexity}

Traditionally, computer and communication networks have been modeled through analytical-based networking. This method of modeling requires a network to be captured entirely by a set of system equations. As networks continue to develop and advance, this notion of exact analytical modeling becomes infeasible due to the inherent complexity of the networks being modeled. 5G networks are a prime example that illustrates this concept as each node is projected to possess over 2000 configurable parameters \cite{EE1}. To mitigate the need for analytical modeling, data-based networking, a paradigm-shifting method of modeling the system, has been suggested. This method of networking does not require an exact system model, but rather it creates a system model out of the available data. When considering ML applications in networking, the data-based networking paradigm is essential. \par
Using data-based networking to model NFV-enabled networks requires incredible amounts of network-generated data. As network complexity increases, so does the size of the network-generated data. This increase in data coupled with the massive variety of the data due to the multitude of configurable parameters available in future networks and the requirement to rapidly process all of this data suggests that is satisfies the high volume, variety, and variability characteristics of “Big Data.” The impact of big data is especially noticeable when considering data-driven network solutions as any update to the network state will result in a significant amount of network generated data that must be processed by the models to ensure their long-term performance. As with any Big Data application, special consideration must be made to acquire, process, analyze, and store this data.

\subsection{Model Scalability and Transferability}

Model scalability is a challenge due to the dynamic nature of networks, and the inability to adapt to severe network changes effectively. New methods of applying ML must be investigated and implemented to mitigate this limitation in scalability. Intelligence has been highlighted as a critical component of 5G and beyond systems; while traditional ML can be used for many problems such as traffic forecasting and NFV MANO problems in static network structures, it is ineffective against a changing domain. As such, ML models capable of adapting to statistical changes in their domain are essential. \par

Another important consideration is the ability of a model to be transferred from one domain to another. NSPs will be operating hundreds of networks spread out across large geographical areas. Ideally, in the case of a failure, intelligence should be quickly re-injected into MANO entities to ensure service continuity and downtime avoidance. By operating many networks simultaneously, NSPs are in a unique position to implement decentralized group learning strategies, which will enable the rapid re-deployment of models in case of failure. 

\subsection{Privacy and Ethical Considerations}

With large corporations worldwide facing severe criticism for their data privacy and security, extreme caution must be taken when developing ML models and systems due to the nature of the data used. In terms of data privacy and security, data governance is paramount when considering that network data can include personal and sensitive information such as personal identifiers and activities. Additionally, since many critical applications, including financial and emergency services, utilize the network, severe measures must be implemented to ensure their protection and regular performance in accordance with QoS guarantees and SLA requirements. However, these are not the only challenges that must be considered when implementing ML models. Increasingly, questions relating to the ethics behind ML models and the data used to train them are being raised. The European Union has created a special task force aimed at the regulation of ML and AI, specifically the ethical considerations behind its implementation \cite{FF1}. Additional governing bodies will likely follow suit and also adopt similar guidelines; therefore, it is essential to continually be aware of the current regulatory guidelines and best practices in the field.

\subsection{Data Acquisition, Processing, Analysis, and Storage}

Data generated from future networks will be classified as Big Data due to its high volume, velocity, and variety. To fully harness the potential of this data, specific processes must be created to ensure that its acquisition, processing, analysis, and storage are efficiently completed given the nature of the data. In terms of the acquisition, care must be taken in determining what data is necessary and what data is redundant. \par
Regarding the processing and analysis of the data, several considerations must be made. Firstly, the method of processing must be defined (\textit {i.e.}, local, cloud, hybrid, distributed). The various processing and analysis schemes must be evaluated against the problem and the objectives; for instance, in an IIoT scenario, to meet ultra-low latency requirements, significant data processing and analysis must be completed at the edge. This requirement inherently adds additional computational requirements to the IIoT network nodes. Additionally, the tradeoffs between the various schemes regarding security and cost must be extensively assessed. While using the cloud can increase the available computing resource, it can also weaken security and increase the overall budget. Selecting the appropriate method to process and analyze the data is vital to the feasibility and successful implementation of the system. \par
The storage of data is also a pressing challenge regarding ML in NFV-enabled networks. Due to the volume and velocity of the data, effective methods of storage must be selected. There are several possible storage schemes, including local, hybrid, and cloud; however, new schemes must be adapted to optimize data availability. This concept is especially important when considering distributed node machine learning techniques whereby ML models are trained and updated on edge and fog nodes, which require storage and compute capacity. Optimizing the storage methods for these nodes is especially vital as they are generally lightweight and have significantly less resource capacity than the traditional core nodes. Furthermore, concerning the topic of model updates and storage, the quantity of update data required, the extent of historical data, and the number of previously trained models must be considered. \par
Traditionally, the availability of large amounts of historical data has been positively regarded; however, with the growing size and faster speed of the generated data, this perception is beginning to change. Concept Drift and Bonferroni’s Principle are two primary data science observations that occur most commonly in ML with Big Data applications. Concept Drift is the idea that over time, the properties of the predicted variables will change, thereby rendering a model ineffective due to a shifting domain. As such, effective update schemes must be created, and an appropriate amount of recent and historical data must be used to optimize the model’s performance and enable its continual success. Bonferroni’s principle states that as the volume of data increases, the observation of spontaneous and unimportant events also increases. This principle is especially concerning as individual relationships might be observed in the data which are unfounded and meaningless with the potential of tampering or misguiding company operations. As such, the quantity of historical data stored must be conducted in a manner that ensures that there is enough data to address Concept Drift but also ensure that the volume of data does not enable misguided observations regarding the Bonferroni Principle. An ML model capable of processing data without the need for excessive storage and communication requirements would be an ideal candidate to mitigate this challenge.

\section{Reinforcement Learning}

RL is the first Advanced AI strategy that will be considered. A description of its mechanics, the benefits associated with it, as well as its potential use cases within the NFV MANO framework will be examined.

\subsection{What is Reinforcement Learning?}

RL is an advanced ML technique that enables the learning of policy decisions in complex environments \cite{GG1}. Through trial and error, the RL agent interacts with its environment and selects an action and receives a reward for the action taken; the higher the reward, the better the action taken. The ultimate goal of the RL agent is to maximize the reward received through the learning of the policies leading to the maximum reward. As environment complexity increases, so does the training time for RL agents; this relationship hindered the adoption of RL in NFV-enabled networks. Deep Reinforcement Learning (DRL) has been proposed to mitigate this drawback, where the rewards of all actions are estimated and learned from a reduced number of training iterations \cite{HH1}. Using the predictive capabilities of Deep Learning (DL) combined with the policy learning of RL, DRL is poised to assume a significant role in NFV MANO. Figure 1 outlines both RL and DRL at a high level. As seen through Figure 1 mechanically, both RL and DRL operate the same way; however, the DRL agent is more advanced and contains DL models used to predict future actions based on environmental observations.
\begin{figure}
\centerline{\includegraphics[width=15pc]{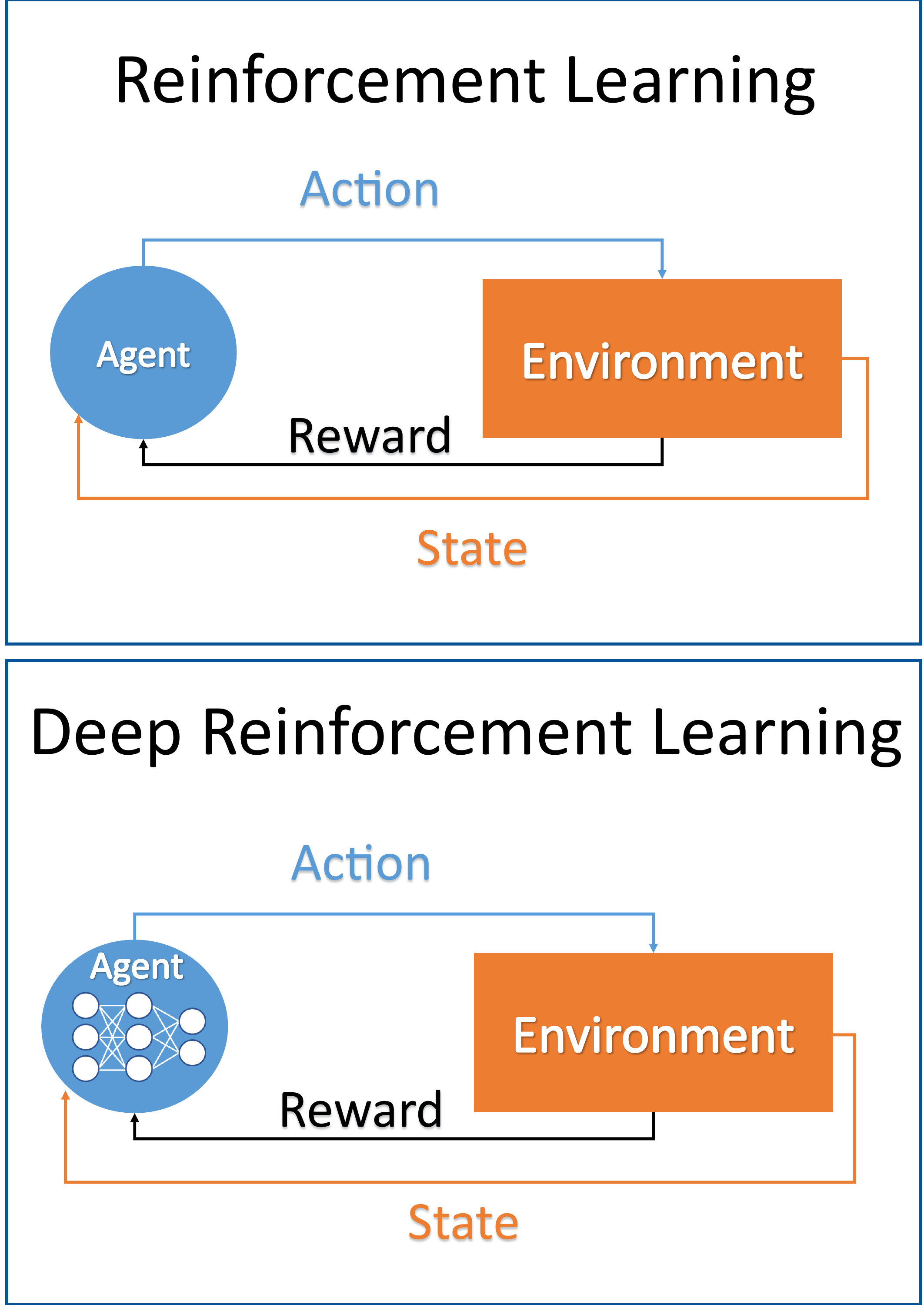}}
\caption{RL vs. DRL}
\end{figure}

\subsection{What are the Benefits of Reinforcement Learning?}

There are several benefits associated with RL, which make it particularly suitable for NFV MANO implementations. Firstly, aside from the setup of the simulation environment and the development of the reward function, human involvement in RL is minimal. Furthermore, RL has the ability to learn strict policy decisions that traditionally would require the formulation of a constrained optimization program, near-optimal heuristic solution, or supervised ML trained on (near)optimal solutions. Perhaps the most significant benefit experienced through RL is the continual adaptability of the agent to a changing domain. Traditional ML can suffer significantly from concept drift (\textit {i.e.}, the real environment changes and is not statistically similar to the training/simulation environment) and require periodic retraining, which is resource consuming, requires human intervention, and can lead to unexpected downtimes. However, as proven through multiple implementations of RL in industries such as video gaming \cite{HH1}, the RL agent is able to adapt to a changing environment and continually learn optimal policies. Another advantage over traditional ML is the rapid rate of transfer learning possible; by transferring an RL agent to a similar domain, the learning of the new domain policies is very time efficient. Another advantage of RL is its ability to perform experiential learning. In contrast with traditional ML approaches, RL does not require a dataset of any kind to learn optimal policies, which inherently give it a significant advantage in situations like NFV Orchestration, where there is a large amount of distributed data that cannot easily be formulated into a single set.

\subsection{How can Reinforcement Learning be used in NFV MANO?}

The ability to learn strict domain policies gives RL the incredible potential to excel at NFV MANO tasks. In short, any task which can traditionally be solved using policy programming and optimization problem formulation can be theoretically solved using RL. Some examples of such problems in NFV MANO include resource allocation, VNF instantiation, VNF placement, VNF migration, VNF scaling, and VNF termination. While many NFV orchestrator functionalities can be achieved in theory using RL, several practical considerations must be addressed. \par
Firstly, the design of the reward function must accurately capture all the policy requirements the RL agent will be required to learn. In some implementations, RL agents have maximized the objective without satisfying the policy requirements due to improper reward function formulation \cite{II1}. Additionally, the training environment needs to be similar to the real environment; therefore, in the case of NFV MANO applications, recent and historical network data must be used for simulation. Additionally, RL agents trained in the same environment might exhibit different performance; therefore, the training of several agents and the selection of the best one will be required. As with traditional ML, RL can suffer from local reward optimum. When considering DRL, the tuning and optimization of the deep network used to predict the rewards must also be considered to optimize performance. Finally, the increasing complexity of networks creates a vast state space for DRL to learn and optimize. Mitigation of the challenges, as mentioned above, presents a multitude of research opportunities and potential specifically when applied to NFV MANO use cases and implementations. 

\section{Federated Learning}
FL is the second Advanced Intelligence strategy that will be considered. A description of its mechanics, the benefits associated with it, as well as its potential use cases within the NFV MANO framework will be examined.

\subsection{What is Federated Learning?}

Federated Learning is an advanced machine learning technique proposed by Google in 2017 \cite{JJ1}. The primary purpose of FL is to provide decentralized collaborative learning across similar domains. In its essence, FL uses a set of nodes with processing capabilities to train local machine learning models and then uses an agent to aggregate all the results into a global model. Looking at the mechanics of FL at a more granular level, there are three main steps involved, global model retrieval, local model training, and global aggregation. Initially, the nodes retrieve the global model from the agent. Once retrieved, the nodes go on to train the global model using their local data. Since each node will have collected different data, through the training process, each node develops a local model. Once a set number of training iterations has been completed, the nodes compare their current local models to the initial global models and provide an update to the aggregation agent. This update, however, only lists the changes made; it does not include the model itself. The aggregator then takes the updates from the nodes and aggregates them into a new model. This process is repeated until a predetermined stopping criterion has been met. Figure 2 highlights the steps mentioned above and provides a high-level overview of FL across nodes A, B, and C. 

\begin{figure}
\centerline{\includegraphics[width=18pc]{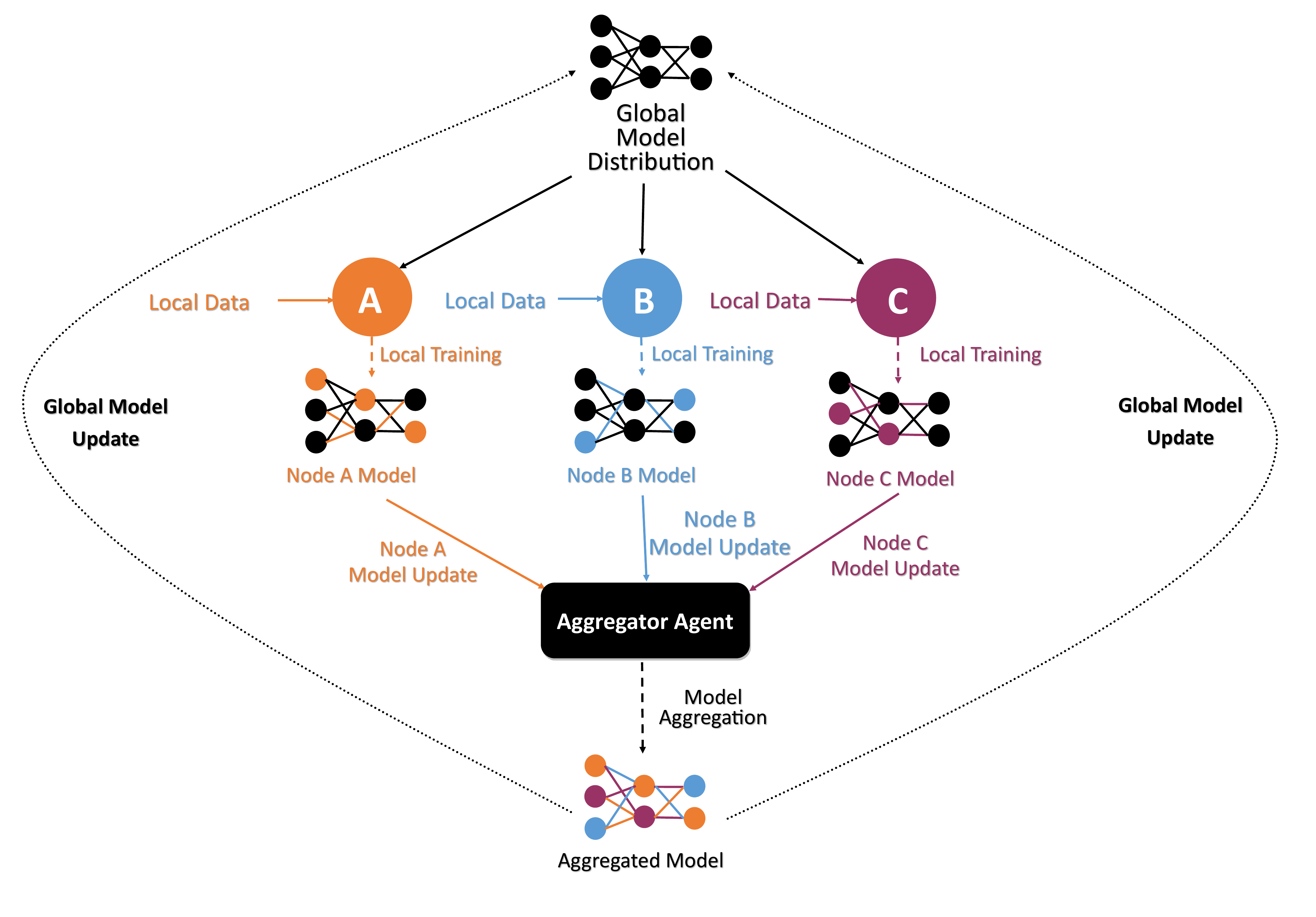}}
\caption{Federated Learning Overview}
\end{figure}

\subsection{What are the Benefits of using Federated Learning?}

The use of FL provides many key improvements upon traditional ML techniques, which make it a prime candidate for use in NFV MANO. Perhaps the most significant advantage of FL is related to its privacy considerations. Since each local node collects and processes its data and simply transmits an update to the aggregator agent, there is no storage of local node data on the cloud, nor is there transferring of local data between nodes. This has the incredible advantage of ensuring local node data remains private while still using insights gained from other nodes to refine the training of the local model further. The ability to handle multiple node’s data to develop a collective insight has enabled the use of FL in situations which traditionally have been hesitant to adopt ML strategies due to privacy concerns; One typical example of this is healthcare due to the sensitivity of the patient’s data \cite{KK1}. Through FL, multiple hospitals have been able to benefit from intelligence insights while still ensuring patient data is private and secure. Additionally, the use of numerous local data sets has enabled an added layer of intelligence since more diverse data can be used to train a single model. Moreover, compared to other distributed learning approaches, FL provides a much more efficient communication scheme since only model updates are being sent to the model aggregator, thereby saving valuable bandwidth and not burdening communication channels. The scheme for model aggregation poses the greatest challenge when creating an FL system as the frequency of updates, and the number of nodes participating in the updates must be configured. Considering a scenario where a federated node loses its local model due to a failure, the global model can be quickly pushed to it and can be used effectively without the need for training; however, through a limited number of training iterations, the local model can regain the level of performance it previously possessed pre-fault. 

\subsection{How can Federated Learning be used in NFV MANO?}

When considering the applicability of FL to NFV MANO, there exists a multitude of possible use cases that can be explored. Simply put, any entity which occurs more than once in a given NSPs NFV MANO and can use intelligence can be a potential candidate for FL. For example, when considering network nodes, data collected by each node can be used to train local models for traffic forecasting, intrusion detection, or fault isolation. Each of these models can be then aggregated into a global model and used in a specific network. If similarities between networks are observed, then nodes across multiple networks can be used to strengthen the model further. Due to the flexibility of NFV technology, FL can be implemented at a very granular level or a very high level. Taking a higher-level approach, NSPs operating data centers can use FL across their datacenters for NFV MANO tasks such as resource estimation and demand forecasting. Considering the highest-level implementation of FL for NFV MANO, NSPs operating different networks can use FL for the training of NFV MANO entities such as the NFV Orchestrator, which is responsible for lifecycle decisions for VNFs such as instantiation, scaling, migration, and termination. Figure 3 illustrates the example outlined above, whereby FL is implemented using local data from an NSP’s geo-distributed datacenters.

\begin{figure}
\centerline{\includegraphics[width=17pc]{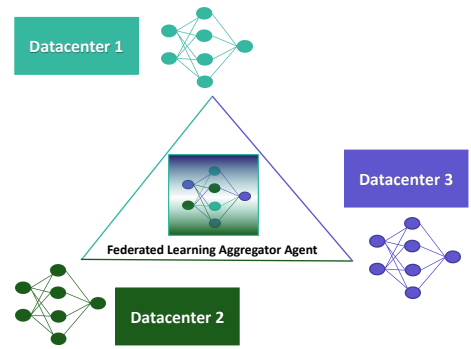}}
\caption{Federated Learning Data Center Use Case}
\end{figure}

\section{Next Steps}

In recent years, considerable research has been done in terms of NFV MANO. Initially, this work considered complex optimization problems and near-optimal heuristic solutions; however, an increasingly significant portion of work is considering the use of ML. Due to the increasing complexity of networks and the incoming introduction of highly configurable 5G technology, leveraging the benefits of ML is paramount. When considering the implementation of ML in NFV MANO, a bottom-up micro-functionality approach should be considered. A comprehensive example of this approach is illustrated through the NFV orchestrator. As mentioned, the main functionalities of the NFV orchestrator include the placement, migration, scaling, and termination of VNFs. The first step considering a bottom-up micro-functionality approach to this entity would require simple ML models for each of its functionalities, as mentioned above, as a proof of concept for ML integration. These simple models would need to perform well compared to leading heuristics, however, would not be expected to fully capture the entire system requirements (i.e., near-optimal vs. optimal). The next step in the process would require the refinement of the simple models and the development of more advanced and sophisticated models to capture the system requirements better and significantly improve upon the performance of near-optimal heuristics and time-infeasible optimal solutions. The third step would require the continual combination and aggregation of these models such that a single ML-based framework which can capture all the functionalities of the NFV orchestrator while adhering to QoS requirements and SLA guarantees. Figure 4 illustrates the outlined bottom-up micro-functionality approach. \par

\begin{figure}
\centerline{\includegraphics[width=17pc]{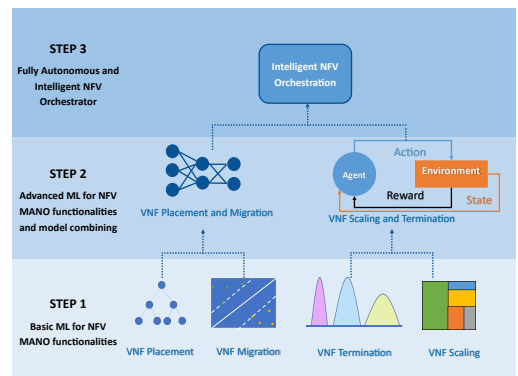}}
\caption{Bottom-Up Micro-Functionality Approach}
\end{figure}

Given the previously outlined approach, RL and FL have a pivotal role to play. Their integration into the second stage of the bottom-up micro-functionality approach is critical. However, these two Advanced Intelligence strategies should not be considered independently. The combination of RL and FL and the adoption of Federated Reinforcement Learning (FRL) is a crucial enabler for this approach and the level of intelligence required for future networks. Some works have already begun implementing FRL \cite{LL1}; however, its application to NFV Orchestration remains an open research challenge. Moving forward, the bottom-up micro-functionality approach combined with FL, RL, and FRL poses an exciting and open research area with significant opportunity for meaningful contributions.

\section{Summary and Conclusions}

The use of advanced intelligence techniques such as RD, DRL, FL, and FRL provides Network Service Providers worldwide with the option of leveraging the benefits of traditional ML techniques while adequately addressing their shortcomings. Figure 5 details the various stages of transition from conventional analytical-based networking to data-based networking and beyond. \par

\begin{figure}
\centerline{\includegraphics[width=17pc]{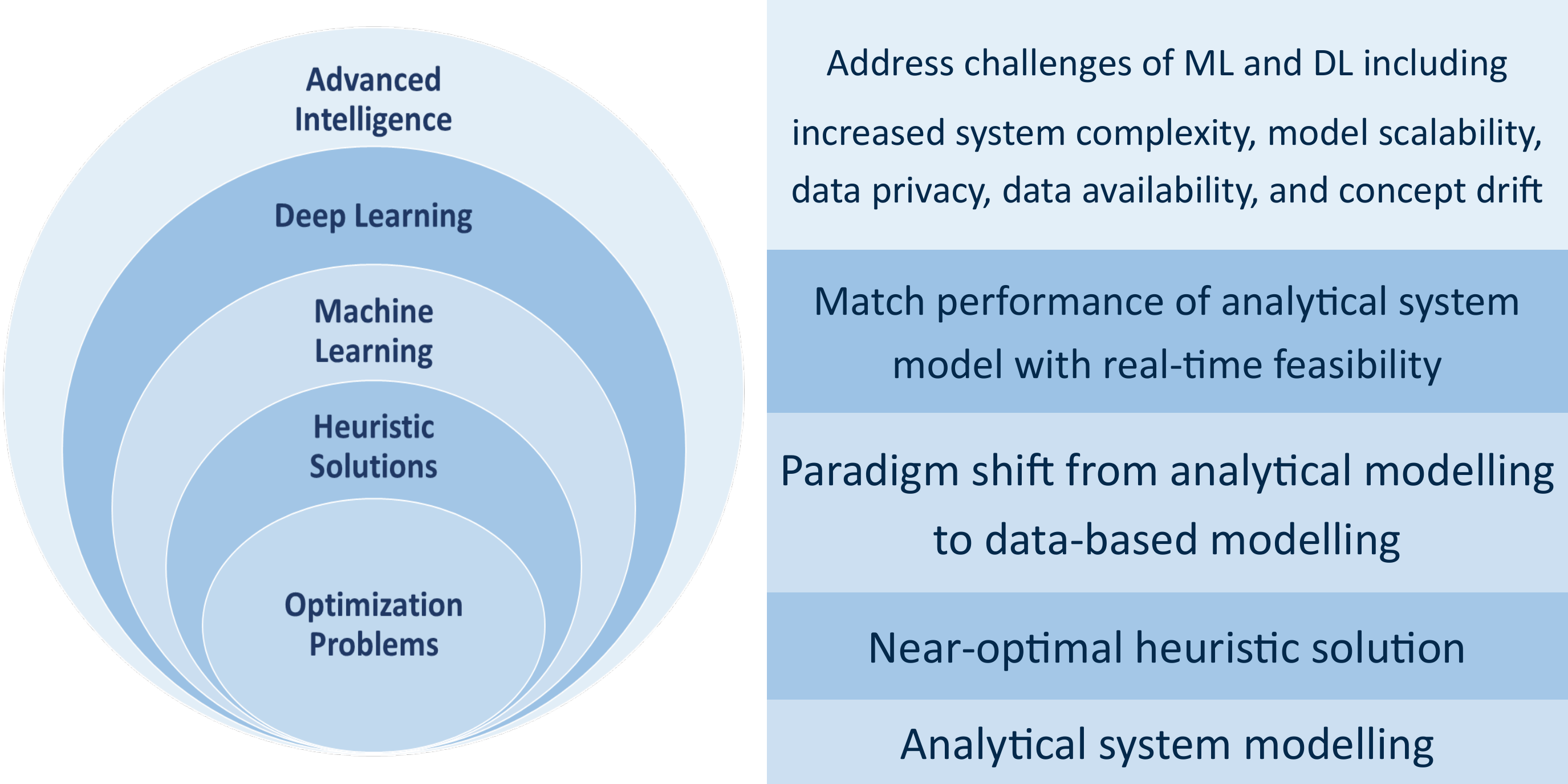}}
\caption{Network Modeling Transition}
\end{figure}

As outlined throughout this paper, both the inherent problem complexity as well as the increasing system complexity create several challenges for NSPs. A summary of these complexity-related challenges is presented in Table 2.
\begin{table}
\caption{Complexity-Related Challenges}
\renewcommand{\arraystretch}{2}
\begin{tabular}{| >{\centering}m{2.5cm}| >{\centering}m{5.5cm}|}
\hline 
Type of Complexity & Summary\tabularnewline
\hline 
\hline 
Inherent Problem Complexity & NP-Hard computational complexity of MANO problems \tabularnewline 
\hline
\multirow{5}{2.5cm}{\centering Increasing System Complexity} & Infeasibility of traditional analytical modelling \tabularnewline 
\cline{2-2} 
& Large volumes of network-generated data requiring storage, processing and analysis classify this problem as a Big Data Application \tabularnewline 
\cline{2-2} 
& Big Data creates a vast observation space for data driven solutions and can impact their training time \tabularnewline 
\hline
\end{tabular}

\end{table}

However, there are still several challenges associated with traditional ML, and the various forms of Advanced Intelligence presented, which must be addressed. Table 2 outlines the various challenges highlighted throughout this paper regarding the use of traditional ML techniques, FL and RL.

\begin{table}
\caption{Challenge Summary}
\renewcommand{\arraystretch}{2}
\begin{tabular}{|c|c|}
\hline 
Technique & Challenges\tabularnewline
\hline 
\hline 
\multirow{4}{*}{ML} & Increasing Network Complexity\tabularnewline 
\cline{2-2} 
 & Model Scalability and Transferability\tabularnewline
\cline{2-2} 
 & Privacy\tabularnewline
\cline{2-2} 
 & Data Acquisition, Processing, Analysis and Storage\tabularnewline
\hline 

\multirow{4}{*}{RL} & Reward Function Design\tabularnewline
\cline{2-2} 
 & Training Environment Simulation\tabularnewline
\cline{2-2} 
 & Model Tuning and Optimization\tabularnewline
\cline{2-2} 
 & Vast Search Space\tabularnewline
\hline 
FL & Model Aggregation Scheme\tabularnewline
\hline 
\end{tabular}

\end{table}

With a plethora of possible use cases for NFV MANO, there are many open research topics and directions that can be pursued. To summarize, Figure 6 lists the benefits of using FL and RL in NFV MANO and provides a list of use cases in NFV MANO. Through the integration of advanced intelligence, NSPs can optimize NFV MANO and prepare for the transition to 5G and beyond networks.
 
\begin{figure}[h]
\centerline{\includegraphics[width=17pc]{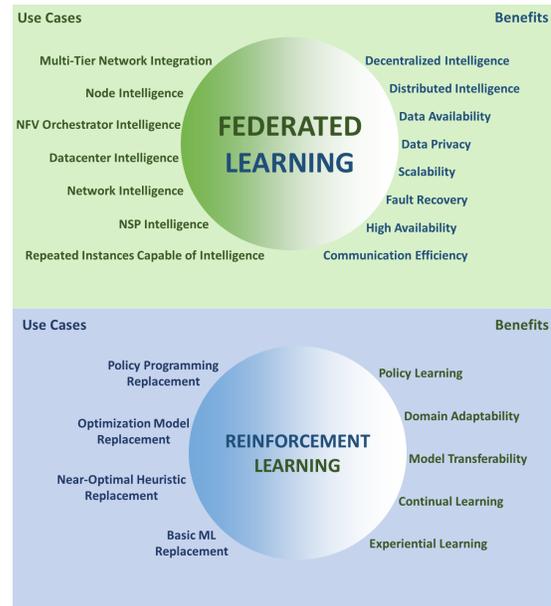}}
\caption{Advanced Intelligence Benefits and Use Cases}
\end{figure}


\begin{thebibliography}{1}

\bibitem{AA1}
ETSI, "Network Functions Virtualisation: An Introduction, Benefits, Enablers, Challenges and Call for Action" . [Online].  Available: \url{ http://portal.etsi.org/NFV/NFV_White_Paper.pdf}

\bibitem{BB1}
 H. Hawilo, A. Shami, M. Mirahmadi and R. Asal, "NFV: state of the art, challenges, and implementation in next-generation mobile networks (vEPC)," {\it IEEE Network}, vol. 28, no. 6, pp. 18-26, 2014.

\bibitem{CC1}
T. S. Buda {\it et al.}, “Can machine learning aid in delivering new use cases and scenarios in 5G?,” {\it Proc. NOMS  IEEE/IFIP Netw. Oper. Manag. Symp.}, pp. 1279–1284,
2016.

\bibitem{RR1}
C. Rostos {\it et al.}, “Network service orchestration standardization: A technology survey,”  {\it Computer Standards \& Interfaces}, vol. 54, pp.203-215, Nov. 2017.

\bibitem{RR2}
J. Gil Herrera and J. F. Botero, "Resource Allocation in NFV: A Comprehensive Survey," in {\it IEEE Transactions on Network and Service Management}, vol. 13, no. 3, pp. 518-532, Sept. 2016.

\bibitem{RR3}
B. Yi, X. Wang, K. Li, S. Das, and M. Huang, “A comprehensive survey of Network Function Virtualization,” in {\it Computer Networks}, vol. 133, pp. 212-262, Mar. 2018.

\bibitem{DD1}
D. M. Manias {\it et al.}, “Machine Learning for Performance-Aware Virtual Network Function Placement,” {\it IEEE GlobeCom}, Waikoloa, USA, 2019, pp. 1-6.

\bibitem{EE1}
M.Wang, Y. Cui, X.Wang, S. Xiao, and J. Jiang, “Machine Learning for Networking: Workflow, Advances and Opportunities,” {\it IEEE Netw.}, vol. 32, no. 2, pp. 92–99, 2018.

\bibitem{FF1}
European Comission, High-Level Expert Group on Artificial Intelligence. [Online]. Available: { https://ec.europa.eu/digital-single-market/en/high-level-expert-group-artificial-intelligence}

\bibitem{GG1}
L. P. Kaelbling, {\it et al.}, "Reinforcement learning: A survey." {\it Journal of artificial intelligence research}, pp.237-285, 1996.

\bibitem{HH1}
V. Mnih, {\it et al.}, "Human-level control through deep reinforcement learning." {\it Nature}, pp. 529-533, 2015.

\bibitem{II1}
Leike, Jan, {\it et al.}, "AI safety gridworlds." {\it arXiv}, preprint, 2017.

\bibitem{JJ1}
J. Konečný, {\it et al.}, "Federated learning: Strategies for improving communication efficiency." {\it arXiv} preprint, 2016.

\bibitem{KK1}
T. S. Brisimi, {\it et al.}, "Federated learning of predictive models from federated electronic health records." {\it International journal of medical informatics} pp. 59-67, 2018.

\bibitem{LL1}
B, Liu, {\it et al.}, "Lifelong federated reinforcement learning: a learning architecture for navigation in cloud robotic systems."{\it  IEEE Robotics and Automation Letters}, pp.4555-4562, 2019.

\end{thebibliography}
\end{document}